# Bounded Satisfiability for PCTL

Nathalie Bertrand[1,2], John Fearnley[2], and Sven Schewe[2]

1   Inria Rennes Bretagne Atlantique, Rennes, France
2   Department of Computer Science, University of Liverpool, Liverpool, UK

**Abstract**

While model checking PCTL for Markov chains is decidable in polynomial-time, the decidability of PCTL satisfiability, as well as its finite model property, are long standing open problems. While general satisfiability is an intriguing challenge from a purely theoretical point of view, we argue that general solutions would not be of interest to practitioners: such solutions could be too big to be implementable or even infinite. Inspired by bounded synthesis techniques, we turn to the more applied problem of seeking models of a bounded size: we restrict our search to implementable – and therefore reasonably simple – models. We propose a procedure to decide whether or not a given PCTL formula has an implementable model by reducing it to an SMT problem. We have implemented our techniques and found that they can be applied to the practical problem of sanity checking – a procedure that allows a system designer to check whether their formula has an unexpectedly small model.

## 1   Introduction

PCTL [9] is a popular logic for the specification of probabilistic systems. The model checking problem for PCTL formulas over Markov chains has been widely studied: it is known to be solvable in polynomial time, and mature tools, such as PRISM [13] and MRMC [10], have been developed. By contrast, satisfiability procedures for PCTL have received much less attention. Recently, it has been shown that the satisfiability problem for the qualitative fragment of PCTL is EXPTIME-complete [4]. However, the satisfiability problem for PCTL itself is not even known to be decidable.

In this paper, we take a step back, and ask the following question: even if we could find an algorithm for the satisfiability of PCTL, would it be useful to a practitioner? We believe that it would not. Even in the qualitative fragment of PCTL, there are already formulas for which there are only infinite state models. Moreover, the problem of deciding, for this fragment of PCTL, whether there is a finite model is EXPTIME-hard [4]. Obviously, the situation is at least as bad for the full PCTL logic.

When practitioners use satisfiability procedures, it is likely that they are interested in whether their formula has an *implementable* model. A constructive satisfiability procedure may return an infinite state model, or a model with bizarre transition probabilities that may be difficult to implement in practice. Neither of these two situations seems to be desirable. Hence, our goal is to solve the following problem.

"Does a PCTL specification $\phi$ have an implementable model?"

Our results are inspired by the work on bounded synthesis for LTL specifications [16, 8, 6, 11, 7], where the question of whether there is a small reactive system for an LTL

This work was supported by the Engineering and Physical Science Research Council grant EP/H046623/1 'Synthesis and Verification in Markov Game Structures' and a Leverhulme Trust Visiting Fellowship.

formula is considered. Building on this, we define the *bounded satisfiability* problem for PCTL formulas, where the goal is to find a *simple* model of a PCTL formula. In our setting, a model is simple if it has a small number of states, and if it uses only rational transition probabilities that can be easily simulated by, for example, coin tossing. We believe that having a simple model is a prerequisite for having an implementable model. Certainly, infinite models, and models whose probabilities cannot be simulated by coin tossing would not seem to be useful in practice.

**Results**

In this paper, we introduce the concept of a simple model, and the bounded satisfiability problem. This is the problem of finding, for a given PCTL formula $\phi$ and bound $b$, a simple model of $\phi$ with at most $b$ states. We provide a reduction from bounded satisfiability to SMT. While PCTL satisfiability is not known to be decidable, our results show that the bounded satisfiability problem can be decided. We also provide complexity results for the bounded satisfiability problem. We show that it is NP-complete in the size of the minimal model. Furthermore, we show that approximating the size of the minimal model is NP-hard.

We have constructed a simple implementation of our reduction from bounded satisfiability to SMT, and we have solved the resulting constraint systems using the Yices SMT solver [5]. We tested this implementation on an academic case study, and our results show that the bounded satisfiability problem can indeed be solved when the number of states required is small.

**Practical Applications**

While our simple implementation does show that small models can be found by our techniques, the size of these models is clearly well below what would be required for constructing systems in an industrial setting. On the other hand, we argue that a bounded synthesis procedure for even a small number of states is useful for the purpose of *sanity checking.*

In model checking, we attempt to verify that a potentially buggy system satisfies a specification. However, in recent years it has become increasingly clear that the specification itself may also contain bugs. See, for example, the work on vacuity checking [3, 15, 12, 1, 2]. It can also be remarkably difficult to detect these errors, since a model checking procedure will simply output "yes" in the case where a buggy system satisfies the buggy specification.

We propose that bounded satisfiability has a role to play in helping system designers find bugs in their specifications. Consider, for instance, a complicated specification of a network protocol that allows, among many other things, the network to go down and subsequently be recovered. A buggy specification may allow a model that goes down immediately after service has been restored, which would allow the model to circumvent most of the specification. This model will have far fewer states than a model that implements a correctly functioning network. In this case, while a correct system may be too large to build with our techniques, it is quite possible that the broken system could be built.

Hence, we propose that bounded satisfiability should be used as a sanity check, in order to test that a formula does not have an error that can be exploited by a small model. Suppose that a system designer has a large system that is known to satisfy some complicated specification. If the bounded satisfiability procedure produces a small model for the specification, then there is a problem that can be resolved in one of two ways. Firstly, it may be the case that the small model does precisely what the designer wants, and in this case the overly-complex large system can be replaced by the small one. The more likely outcome is that the small model does not do what the designer intended. In this case, the designer now has a small, understandable counter-example, which can be used to help correct the specification.

Our experimental results show that our bounded satisfiability procedure is particularly suitable for sanity checking. While our implementation does not seem to scale well with the number of states in the model, it does scales well with size of the input formula. This indicates that sanity checking may indeed be possible for the type formulas that are used in practice.

## 2 Preliminaries

### 2.1 Markov chains

We recall below the definition of discrete-time Markov chains, simply referred to as Markov chains in the sequel.

▶ **Definition 1** (Markov chain). A Markov chain is a tuple $\mathcal{M} = (S, \mathbf{P}, \iota, L)$, where $S$ is a finite or countable set of states, $\mathbf{P} : S \times S \to [0,1]$ is a probabilistic transition function, $\iota \in S$ is an initial state, and $L$ is a labelling function mapping states to atomic propositions. $\mathbf{P}$ satisfies, for all $s \in S$:

$$\sum_{s' \in S} \mathbf{P}(s, s') = 1$$

A *path* in $\mathcal{M}$ is a sequence of states $\pi = s_0 s_1 \cdots \in S^*$ such that, for every $i \in \mathbb{N}$, $\mathbf{P}(s_i, s_{i+1}) > 0$. The set of all paths starting in a state $s \in S$ is denoted $\mathsf{Paths}(s)$. In order to define a probability measure $\Pr_\mathcal{M}$ over suitable sets of paths, we first explain how a measure is associated with basic sets of paths called cylinder sets, which gather all paths sharing a given finite prefix. For $s_0 s_1 \cdots s_k$ a finite sequence of states, we let $\mathsf{Cyl}(s_0 s_1 \cdots s_k) = \{\pi \in \mathsf{Paths}(s_0) \mid s_0 s_1 \cdots s_k \prec \pi\}$ where $\prec$ is the usual prefix order, and define its measure as $\Pr_\mathcal{M}^{s_0}(\mathsf{Cyl}(s_0 s_1 \cdots s_k)) = \prod_{0 \le i < k} \mathbf{P}(s_i, s_{i+1})$. For $s \ne s_0$, $\Pr_\mathcal{M}^s(\mathsf{Cyl}(s_0 s_1 \cdots s_k)) = 0$.

Now $\Pr_\mathcal{M}^s$ can be extended to a set of reasonable sets of paths, namely the $\sigma$-algebra generated from cylinder sets. This $\sigma$-algebra over $\mathsf{Paths}(\mathcal{M}) = \bigcup_{s \in S}(\mathsf{Paths}(s))$ precisely consists of the smallest collection of sets of paths in $\mathcal{M}$ that contains the empty set, all $\mathsf{Cyl}(s_0 s_1 \cdots s_k)$ for any finite sequence $s_0 s_1 \cdots s_k$ of states, and is closed under complementation and countable union. The extension of $Pr_\mathcal{M}^s$ from cylinders to the $\sigma$-algebra they generate is unique, and we still denote it $\Pr_\mathcal{M}^s$. Note that not all sets of paths are measurable with respect to $\Pr_\mathcal{M}^s$, but the sets we will consider in this paper are simple enough to avoid such difficulties. We use $\Pr_\mathcal{M}$ as an abbreviation of $\Pr_\mathcal{M}^\iota$.

### 2.2 PCTL

Probabilistic computation tree logic (PCTL) [9] is a probabilistic variant of CTL, where path quantifiers are replaced by probabilistic operators. PCTL is interpreted over Markov chains, and one can for example specify that the probability measure of the set of paths satisfying a given until property exceeds some threshold. Formally, the syntax of PCTL is given by the following grammar:

▶ **Definition 2** (PCTL syntax). Let $AP$ be a set of atomic propositions. The syntax of a

PCTL formula is:

$$\phi ::= \top \mid a \mid \phi \wedge \phi \mid \neg \phi \mid \mathbb{P}_{\bowtie \lambda} \tau$$
$$\tau ::= \bigcirc \phi \mid \phi \, \mathcal{U} \, \phi \mid \phi \, \mathcal{U}^{\leq n} \, \phi$$

where $a \in AP$ is an atomic proposition, $\bowtie$ is a comparison operator in $\{<, \leq, =, \geq, >\}$, $\lambda \in [0,1]$ is a rational threshold, and $n \in \mathbb{N}$.

Formulas produced by the production rules of $\phi$ and $\tau$ are called state formulas and path formulas, respectively. State formulas are also called PCTL formulas. PCTL formulas are interpreted over Markov chains. The *step-bounded until* operator has the intuitive semantics that $\mathbb{P}_{\bowtie \lambda}(\phi \, \mathcal{U}^{\leq n} \, \psi)$ is true if the probability is $\bowtie \lambda$ that: $\psi$ holds within the next $n$ steps, and $\phi$ is true until $\psi$ is true. Hence, the step-unbounded until formula $\mathbb{P}_{\bowtie \lambda}(\phi \, \mathcal{U} \, \psi)$ can be thought of $\mathbb{P}_{\bowtie \lambda}(\phi \, \mathcal{U}^{\leq \infty} \, \psi)$.

The syntax and semantics of PCTL only differ from those of CTL by using probabilistic path operators $\mathbb{P}_{\bowtie \lambda}(\bigcirc \cdots)$, $\mathbb{P}_{\bowtie \lambda}(\cdots \mathcal{U} \cdots)$ and $\mathbb{P}_{\bowtie \lambda}(\cdots \mathcal{U}^{\leq n} \cdots)$ instead of universal and existential ones.

▶ **Definition 3** (PCTL semantics). Let $\mathcal{M} = (S, \mathbf{P}, \iota, L)$ be a Markov chain, $s \in S$ a state of $\mathcal{M}$, and $\phi, \phi'$ PCTL formulas. We have:
- $\mathcal{M}, s \models a$ iff $a \in L(s)$,
- $\mathcal{M}, s \models \phi \wedge \phi'$ iff $\mathcal{M}, s \models \phi$ and $\mathcal{M}, s \models \phi'$,
- $\mathcal{M}, s \models \neg \phi$ iff $\mathcal{M}, s \not\models \phi$,
- $\mathcal{M}, s \models \mathbb{P}_{\bowtie \lambda} \tau$ iff $\Pr^s_{\mathcal{M}}(\{\pi \in \mathsf{Paths}(s) \mid \pi \models \tau\}) \bowtie \lambda$.

Finally, for a path $\pi = s_0 s_1 \cdots \in \mathsf{Paths}(s)$ and PCTL formulas $\phi, \phi'$ we have:
- $\mathcal{M}, \pi \models \bigcirc \phi$ iff $s_1 \models \phi$,
- $\mathcal{M}, \pi \models \phi \, \mathcal{U} \, \phi'$ iff $\exists i.\ s_i \models \phi'$ and $\forall j < i.\ s_j \models \phi$,
- $\mathcal{M}, \pi \models \phi \, \mathcal{U}^{\leq n} \, \phi'$ iff $\exists i \leq n.\ s_i \models \phi'$ and $\forall j < i.\ s_j \models \phi$.

Note that the semantics is well-defined because specified sets of paths are indeed measurable. We use the usual shorthand notations known from CTL, such as $\Diamond \, \phi \equiv \top \, \mathcal{U} \, \phi$ and $\Box \, \phi \equiv \neg \Diamond \, \neg \phi$. Using duality of eventually and always operators and the duality of lower and upper bounds of our probabilistic path operators, we can, for example, express $\mathbb{P}_{\leq \lambda}(\Box \, \phi) \equiv \mathbb{P}_{\geq 1-\lambda}(\Diamond \, \neg \phi)$.

## 3  Setting and problem statement

The model checking problem for PCTL over Markov chains is known to be solvable in polynomial time [9]. In contrast to this, satisfiability, that is, the decision problem that asks whether or not a formula has a model, is a long standing open problem for PCTL. Only recently, satisfiability for the restricted fragment of qualitative PCTL, where thresholds can only take value 0 or 1, has been shown to be decidable (EXPTIME-complete [4]). Already this qualitative fragment does not have the finite model property. As an example formula $\phi = \mathbb{P}_{>0}\big(\Box \, (\neg a \wedge \mathbb{P}_{>0} \bigcirc a)\big)$ is satisfiable (it has a model with infinitely many states) but has no finite model [4]. Whether or not a formula has a model surely is a challenging theoretical question, but ultimately not a question of practical interest, especially if all of its models are infinite. On the contrary, an interesting question in practice is whether or not a formula has a *simple* model, where by 'simple' one intends reasonably small and implementable, at least. The problem is then to determine whether or not a formula admits a model with a bounded number of states. This bounded satisfiability problem [16, 8, 6, 11, 7] has first been studied in [16] for LTL and distributed architectures, where it is reduced to an SMT problem.

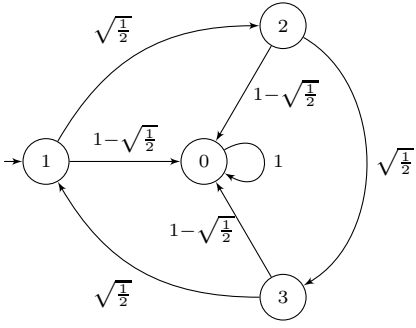
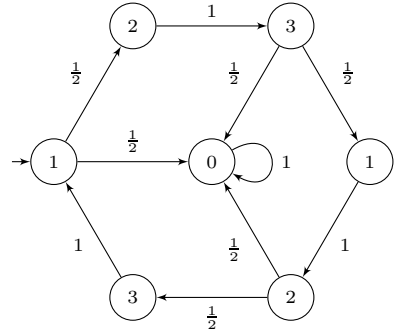

**Figure 1** An irrational model for $\psi_0$.

**Figure 2** A simple model for $\psi_0$.

In the context of the probabilistic branching time logic PCTL and Markov chain models, we advocate that non-implementability can not only result from a large (or even infinite) state space, but also from the transition probabilities. The representation of arbitrary rational or irrational probabilities definitely forms a source of complexity when it comes to implementing a system described by a model. To illustrate this problem, let us consider the following PCTL formula over 2 atomic propositions $\{p, q\}$, which are used to encode $0 \equiv \neg p \wedge \neg q$, $1 \equiv \neg p \wedge q$, $2 \equiv p \wedge \neg q$, and $3 \equiv p \wedge q$.

$$\psi_0 := 1 \wedge \mathbb{P}_{=1}\square \left( \Big(0 \to \mathbb{P}_{=1}(\bigcirc 0)\Big) \right.$$
$$\wedge \Big(1 \to \mathbb{P}_{=1}(\bigcirc (2 \vee 0))\Big) \wedge \Big(2 \to \mathbb{P}_{=1}(\bigcirc (3 \vee 0))\Big) \wedge \Big(3 \to \mathbb{P}_{=1}(\bigcirc (1 \vee 0))\Big)$$
$$\left. \wedge \Big(1 \to \mathbb{P}_{=1/2}(1 \vee 2\, \mathcal{U}\, 3)\Big) \wedge \Big(2 \to \mathbb{P}_{=1/2}(2 \vee 3\, \mathcal{U}\, 1)\Big) \wedge \Big(3 \to \mathbb{P}_{=1/2}(3 \vee 1\, \mathcal{U}\, 2)\Big) \right)$$

The formula specifies that the initial state is labelled with 1, and the 0 states form a sink. It also requires that all successors of states labelled 1 are labelled 2 or 0 (and similarly for 2 and 3); and last, with probability exactly $\frac{1}{2}$, from the state labelled 1, only 1 or 2 are visited until the state labelled 3 is reached (symmetrically for 2 and 3). A model with four states for this formula is represented in Figure 1. Although the bounds in $\psi_0$ are all rationals (as required by the PCTL syntax), the formula forces every four-state model to have irrational probabilities on their edges. Indeed, the model shown in Figure 1 is the only four state model of $\psi_0$. The first two lines of $\psi_0$, for example, require that the initial state is labelled with one, and that all successors of states labelled with 0 must also be labelled with 0. Also, successors of state 1 are among 0 and 2, and so forth. Moreover, letting $\mathbf{P}(i, j)$ be the probability to transition from state $i$ to state $j$, the three last conjuncts in $\psi_0$ imply that $\mathbf{P}(1, 2) \cdot \mathbf{P}(2, 3) = \mathbf{P}(2, 3) \cdot \mathbf{P}(3, 1) = \mathbf{P}(3, 1) \cdot \mathbf{P}(1, 2) = \frac{1}{2}$. This yields $\mathbf{P}(1, 2) = \mathbf{P}(2, 3) = \mathbf{P}(3, 1) = \sqrt{\frac{1}{2}}$. We emphasise that models with irrational probabilities are not implementable, because they require infinite memory.

If we use the same example formula $\psi_0$ but allow for a larger number of states, then the specification becomes easier to satisfy. Seven states suffice for the rational model shown in Figure 2. This model uses only rational probabilities. Moreover, all transitions carry the probability $\frac{1}{2}$ or 1. If we allow for multi-graphs, then we can split transitions with probabilities 1 into two transitions with probability $\frac{1}{2}$ with the same source and target. This inspires our definition of simple models: this example model can be represented as a multi-graph with two (potentially equivalent) outgoing transitions per node, each of which is taken with probability $\frac{1}{2}$.

Moreover, any rational probability $\frac{p}{q}$ can be written using its binary encoding of the form $0.vw^\omega$ with $v, w \in \{0, 1\}^*$. A transition from state $s$ to state $s'$ with probability $\frac{3}{10}$, for example, can be written $0.0(1001)^\omega$, and can be encoded by the Markov chain below, where all transition probabilities are $\frac{1}{2}$:

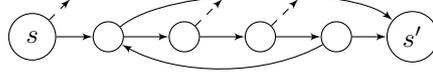

We now introduce *simple Markov chains* (SMCs) and discuss the semantics of PCTL when applied to simple Markov chains.

▶ **Definition 4** (Simple Markov Chains). A Markov chain $\mathcal{M} = (S, \mathbf{P}, \iota, L)$ is called a *simple Markov chain* (SMC) if it satisfies the following.
- The state space of $\mathcal{M}$ is finite ($|S| \in \mathbb{N}$).
- The domain of the probability function $\mathbf{P}$ is $\{0, \frac{1}{2}, 1\}$.
- $\mathcal{M}$ contains an atomic proposition $p_\exists$ that cannot be used in specifications. A state $s$ is called a *real* state if $p_\exists$ is contained in its label ($p_\exists \in L(s)$), and it is called a *hidden* state otherwise.
- The initial state $\iota$ is real, and, from each state, there is a path to a real state.

Intuitively, hidden states are the states used to simulate probabilities that are not $\frac{1}{2}$, and the last constraint guarantees that the probability measure of paths that eventually always stay in hidden states is 0. For the purposes of PCTL, the hidden states should not be counted towards the truth of a path formula. In particular, the next operator should refer to the next real state, and the bounded until operator with bound $n$ should refer to an until that should be satisfied after $n$ real states have been seen. The SMC semantics thus differs from the standard PCTL semantics in the definition of path formulas, which we now redefine.

▶ **Definition 5** (SMC Semantics). The semantics of state formulas is as in Definition 3, while the definition of path formulas changes as follows:
- $\mathcal{M}, \pi \models_{SMC} \bigcirc \phi$ iff $\exists i.\ s_i \models_{SMC} \phi$ and $s_i$ is real and $\forall 0 < j < i.\ s_j$ is hidden,
- $\mathcal{M}, \pi \models_{SMC} \phi\, \mathcal{U}\, \phi'$ iff $\exists i.\ s_i \models_{SMC} \phi'$ and $s_i$ is real and $\forall j < i$. either $s_j \models_{SMC} \phi$ or $s_j$ is hidden,
- $\mathcal{M}, \pi \models_{SMC} \phi\, \mathcal{U}^{\leq n}\, \phi'$ iff $\exists i.\ s_i \models_{SMC} \phi'$ and $s_i$ is real and $\forall j < i$. either $s_j \models_{SMC} \phi$ or $s_j$ is hidden and $|\{j < i \mid s_j \text{ is real }\}| \leq n$.

Given a Markov chain with rational transition probabilities, we have already shown how an equivalent simple Markov chain can be constructed, by simulating the transition probabilities with hidden states. On the other hand, every simple Markov chain has an equivalent Markov chain, which can be obtained by computing the probability of moving from one real state to the next in the simple Markov chain. Therefore, we obviously have the following equivalence.

▶ **Proposition 6.** *For a PCTL formula $\phi$, there is a finite Markov chain $\mathcal{M}$ with rational transition probabilities and $\mathcal{M} \models \phi$ if, and only if, there is a simple Markov chain $\mathcal{M}'$ that satisfies $\mathcal{M}' \models_{SMC} \phi$.*

Having only transition probabilities $\frac{1}{2}$ and 1 is very convenient in practice, because then all random choices can immediately be simulated by a device as simple as tossing a fair coin. The example formula $\psi_0$ and the argumentation above motivate the quest for models that are simple in two respects: they have a reasonable number of states, and all transition probabilities are equal to $\frac{1}{2}$ or 1. We can now formally state our bounded satisfiability problem:

▶ **Definition 7** (Bounded satisfiability problem). **Input:** A PCTL formula $\phi$ and a bound $b \in \mathbb{N}$.
**Question:** Does there exist a simple Markov chain $\mathcal{M}$ with at most $b$ states, such that $\mathcal{M} \models_{SMC} \phi$?

For PCTL, the synthesis problem can be solved by a satisfiability algorithm. Suppose we want a probabilistic environment that gives us proposition p with probability 0.5, and not p with probability 0.5. Then, we can add the following requirement to our formula:

$$\mathbb{P}_{=1}\square\big(\mathbb{P}_{=0.5}(\bigcirc p) \wedge \mathbb{P}_{=0.5}(\bigcirc \neg p)\big).$$

This can obviously be generalised for multiple input propositions, and non-uniform probability distributions.

To check whether there is a simple model $\mathcal{M}$ of $\phi$ with $b$ states that satisfies a formula can clearly be done in non-deterministic polynomial time in the size of the related model checking problem (that is, in the joint size of the model and specification): it can simply guess the model and then check its correctness in polynomial time. It is also NP hard.

▶ **Proposition 8.** *For a given PCTL formula $\phi$ and a bound $b$, the problem of deciding whether there is a simple model $\mathcal{M}$ with $\mathcal{M} \models_{SMC} \phi$ that has $b$ states is NP-complete in the joint size of $\phi$ and $\mathcal{M}$.*

Moreover, we can also show that approximating the size of a smallest model is NP-hard. The reason for this is that it is simple to find a PCTL formula for which the smallest model is of size $\approx 2^n$, where $n$ is polynomial in the specification. Hence, we can construct a PCTL formula $\phi$ that either requires that a boolean formula $\psi$ is true in the first state, or requires that we build an exponential model. Therefore, if $\psi$ has a satisfying assignment, then $\phi$ has a model of size 1, otherwise the smallest model of $\phi$ has exponentially many sates.

▶ **Proposition 9.** *It is NP-hard to approximate the size of the smallest model of a PCTL formula $\phi$ within a factor that is polynomial in $|\phi|$.*

## 4 Reduction to an SMT problem

A satisfiability modulo theories (SMT) problem is the decision problem that consists in determining whether or not a logical formula expressed in boolean logic and using additional theories is satisfiable. Let $\phi$ be a PCTL formula over a set of atomic propositions $AP$. Suppose that we wish to solve the bounded satisfiability problem for $\phi$ with the bound $b$. In this section we will construct a system of SMT constraints that are satisfiable if, and only if, the formula $\phi$ has a model with $b$ states. The size of the SMT formula will be linear in the number of sub-formulas in $\phi$, and the SMT formula can be constructed in linear time. The theories we use in our SMT constraints are linear real arithmetic and uninterpreted function symbols.

### 4.1 The model

We begin by introducing the functions and constraints that will define our model. We define the type $\mathsf{States} = \{1, 2, \ldots, b\}$, with the intention that each integer in $\mathsf{States}$ will represent one state of the model. We define the following functions.

- We define the *left* successor function $\mathsf{left} : \mathsf{States} \to \mathsf{States}$ and the *right* successor function $\mathsf{right} : \mathsf{States} \to \mathsf{States}$. These functions give the two outgoing transitions from each of the states. A transition from a state $s$ to a state $s'$ with probability 1 can be simulated by setting $\mathsf{left}(s) = \mathsf{right}(s) = s'$.
- We define the *existence* function $\mathsf{exists} : \mathsf{States} \to \mathbb{B}$, where $\mathsf{exists}(s)$ is true if $s$ is a real state, and false if $s$ is a hidden state.
- For each atomic proposition $a \in AP$, we define a function $\mathsf{truth}_a : \mathsf{States} \to \mathbb{B}$, where $\mathsf{truth}_a(s)$ indicates that the atomic proposition $a$ is true in state $s$.

Using the functions we have defined, it is possible to define a model that visits only a finite number of real states before getting stuck in hidden states. To avoid this, we introduce a function $\mathsf{dist}_\exists : \mathsf{States} \to [0, 1]$, and the following two constraints:

$$\forall s \cdot \mathsf{exists}(s) \leftrightarrow \mathsf{dist}_\exists(s) = 0,$$

$$\forall s \cdot \neg\mathsf{exists}(s) \to \Big(\mathsf{dist}_\exists(s) > \mathsf{dist}_\exists(\mathsf{left}(s))\Big) \vee \Big(\mathsf{dist}_\exists(s) > \mathsf{dist}_\exists(\mathsf{right}(s))\Big).$$

The first constraint states that $\mathsf{dist}_\exists(s)$ may only be 0 when $s$ is real. The second constraint states that at each hidden state, the value of $\mathsf{dist}_\exists(s)$ must be strictly larger than either $\mathsf{dist}_\exists(\mathsf{left}(s))$ or $\mathsf{dist}_\exists(\mathsf{right}(s))$.

We argue that, if the model satisfies these two constraints, then it cannot get stuck in hidden states. A hidden state $s$ can satisfy the second constraint if, and only if, there is a finite path from $s$ to a real state. Hence, there is some probability $p$, with $p > 0$, to move from $s$ to some real state. Since this holds for all hidden states, the probability of not eventually reaching a real state must be 0.

### 4.2 The formula

For each sub-formula $\psi$ of $\phi$, we associate a function $\mathsf{sat}_\psi : \mathsf{States} \to \mathbb{B}$, where $\mathsf{sat}_\psi(s)$ will be true if and only if state $s$ satisfies $\psi$. In this section we will describe the constraints that are placed on $\mathsf{sat}_\psi$.

#### Non-temporal operators

We begin by giving the constraints for $\mathsf{sat}_\psi$ for the case where $\psi$ is a non-temporal operator. We define the following constraints:

- If $\psi$ is an atomic proposition $a \in AP$, then we add the constraint:

$$\forall s \cdot \mathsf{sat}_\psi(s) \leftrightarrow \mathsf{truth}_a(s). \tag{1}$$

- If $\psi$ is $\neg\psi'$, then we add the constraint:

$$\forall s \cdot \mathsf{sat}_\psi(s) \leftrightarrow \neg\mathsf{sat}_{\psi'}(s). \tag{2}$$

- If $\psi$ is $\psi_1 \wedge \psi_2$, then we add the constraint:

$$\forall s \cdot \mathsf{sat}_\psi(s) \leftrightarrow \mathsf{sat}_{\psi_1}(s) \wedge \mathsf{sat}_{\psi_2}(s). \tag{3}$$

**Next formulas**

We now define the constraints on $\mathsf{sat}_\psi(s)$ in the case where $\psi$ is $\mathbb{P}_{\bowtie\lambda}(\bigcirc \psi')$. It should be noted that we are only interested in whether $\psi'$ holds in the next real state, and that we must account for the fact that several hidden states may be visited before we arrive at a real state. It is for this reason that we introduce the function $\mathsf{value}_\psi : \mathsf{States} \to [0,1]$. Our intention is that this function should give, for each hidden state, the probability that $\psi'$ holds in the next real state. We define the following constraints on $\mathsf{value}_\psi(s)$.

$$\forall s \cdot \mathsf{exists}(s) \wedge \mathsf{sat}_{\psi'}(s) \to \mathsf{value}_\psi(s) = 1,$$
$$\forall s \cdot \mathsf{exists}(s) \wedge \neg\mathsf{sat}_{\psi'}(s) \to \mathsf{value}_\psi(s) = 0,$$
$$\forall s \cdot \neg\mathsf{exists}(s) \to \mathsf{value}_\psi(s) = \frac{1}{2} \cdot \Big(\mathsf{value}_\psi(\mathsf{left}(s)) + \mathsf{value}_\psi(\mathsf{right}(s))\Big).$$

The first two constraints set the value of a real state $s$ to be either 1 or 0 depending on whether $\psi'$ holds at $s$. The final constraint sets the value of a hidden state to be the average of the values of its successors. Recall that we have already introduced constraints to ensure that the system cannot get stuck in hidden states. Therefore, these constraints are sufficient to force, for each hidden state $s$, the function $\mathsf{value}_\psi(s)$ to give the probability that $\psi'$ holds at the next real state.

We can now introduce the constraint for $\mathsf{sat}_\psi(s)$. This constraint simply checks whether the probability of $\psi'$ occurring in the next real state satisfies the bound $\bowtie \lambda$.

$$\forall s \cdot \mathsf{sat}_\psi(s) \leftrightarrow \frac{1}{2} \cdot \Big(\mathsf{value}_\psi(\mathsf{left}(s)) + \mathsf{value}_\psi(\mathsf{right}(s))\Big) \bowtie \lambda.$$

**Until formulas**

We now define the constraints on $\mathsf{sat}_\psi(s)$ in the case where $\psi$ is $\mathbb{P}_{\bowtie\lambda}(\psi_1 \,\mathcal{U}\, \psi_2)$. Following our approach for the next operator, we once again define a function $\mathsf{value}_\psi : \mathsf{States} \to [0,1]$. Our intention is that $\mathsf{value}_\psi(s)$ should give the probability that the until formula $\psi$ is satisfied at the state $s$. We place the following constraints on the function $\mathsf{value}_\psi$:

$$\forall s \cdot \mathsf{exists}(s) \wedge \mathsf{sat}_{\psi_2}(s) \to \mathsf{value}_\psi(s) = 1,$$
$$\forall s \cdot \mathsf{exists}(s) \wedge \neg\mathsf{sat}_{\psi_1}(s) \wedge \neg\mathsf{sat}_{\psi_2}(s) \to \mathsf{value}_\psi(s) = 0,$$
$$\forall s \cdot \neg\mathsf{exists}(s) \vee \big(\mathsf{sat}_{\psi_1}(s) \wedge \neg\mathsf{sat}_{\psi_2}(s)\big) \to$$
$$\mathsf{value}_\psi(s) = \frac{1}{2} \cdot (\mathsf{value}_\psi(\mathsf{left}(s)) + \mathsf{value}_\psi(\mathsf{right}(s))).$$

The first constraint sets the probability to 1 for real states that satisfy the right-hand side of the until, and the second constraint sets the probability to 0 for real states that satisfy neither the left-hand side nor the right-hand side of the until. The final constraint deals with real states that only satisfy the left-hand side of the until, and with hidden states. In both of these cases, we set the probability of the state to be the average of the probability of its successors.

In contrast to the next operator, the constraints that we have introduced are not sufficient to capture the probability of an until operator. To see the problem, consider a model in which $\psi_1$ is satisfied at all states, and $\psi_2$ is not satisfied at any state. Our constraints so far would allow $\mathsf{value}_\psi(s)$ to take any value in $[0,1]$ in such a model. To solve this, we introduce a distance function $\mathsf{dist}_\psi : \mathsf{States} \to [0,1]$, which ensures that $\psi_2$ can be reached with non-zero probability.

$$\forall s \cdot \mathsf{exists}(s) \wedge \mathsf{sat}_{\psi_2}(s) \leftrightarrow \mathsf{dist}_\psi(s) = 0,$$
$$\forall s \cdot \mathsf{value}_\psi(s) = 0 \leftrightarrow \mathsf{dist}_\psi(s) = 1,$$
$$\forall s \cdot \mathsf{value}_\psi(s) \neq 0 \wedge \big(\neg\mathsf{exists}(s) \vee \neg\mathsf{sat}_{\psi_1}(s)\big) \to$$
$$\big(\mathsf{dist}_\psi(s) > \mathsf{dist}_\psi(\mathsf{left}(s))\big) \vee \big(\mathsf{dist}_\psi(s) > \mathsf{dist}_\psi(\mathsf{right}(s))\big).$$

If a state satisfying $\psi_2$ can be reached with non-zero probability from a state $s$, then it is clear that we can set $\mathsf{dist}_\psi(s) < 1$. On the other hand, if no state satisfying $\psi_2$ can be reached from $s$, then the third constraint cannot be satisfied. Therefore, the second constraint must be used, which sets $\mathsf{dist}_\psi(s) = 1$, and then correctly sets $\mathsf{value}_\psi(s) = 0$.

Having specified the value function $\mathsf{value}_\psi$, the constraint for the function $\mathsf{sat}_\psi$ simply compares the value to the bound given by $\lambda$:

$$\forall s \cdot \mathsf{exists}(s) \wedge \mathsf{sat}_\psi(s) \leftrightarrow \mathsf{value}_\psi(s) \bowtie \lambda.$$

**Bounded until formulas**

We now give the constraints for $\mathsf{sat}_\psi$ for the case where $\psi$ is $\mathbb{P}_{\bowtie\lambda}(\psi_1 \, \mathcal{U}^{\leq n} \, \psi_2)$. The constraints that we introduce here can be seen as a generalisation of the constraints that are used for next formulas. For each $i$ in the range $0 \leq i \leq n$, we introduce a function $\mathsf{value}_{\psi,i} : \mathsf{States} \to [0,1]$. The function $\mathsf{value}_{\psi,i}(s)$ is intended to give the probability that $\psi_1 \, \mathcal{U}^{\leq i} \, \psi_2$ holds at the state $s$. We can start by giving a constraint for the function $\mathsf{value}_{\psi,0}$:

$$\forall s \cdot \mathsf{value}_{\psi,0}(s) = 1 \leftrightarrow \mathsf{exists}(s) \wedge \mathsf{sat}_{\psi_2}(s),$$
$$\forall s \cdot \mathsf{value}_{\psi,0}(s) = 0 \leftrightarrow \neg\big(\mathsf{exists}(s) \wedge \mathsf{sat}_{\psi_2}(s)\big).$$

Having defined $\mathsf{value}_{\psi,0}$, we can now define $\mathsf{value}_{\psi,i}$ inductively. For each $i$ in the range $1 \leq i \leq n$, we add the constraints:

$$\forall s \cdot \mathsf{exists}(s) \wedge \mathsf{sat}_{\psi_2}(s) \to \mathsf{value}_{\psi,i}(s) = 1,$$
$$\forall s \cdot \mathsf{exists}(s) \wedge \neg\mathsf{sat}_{\psi_1}(s) \wedge \neg\mathsf{sat}_{\psi_2}(s) \to \mathsf{value}_{\psi,i}(s) = 0,$$
$$\forall s \cdot \mathsf{exists}(s) \wedge \mathsf{sat}_{\psi_1}(s) \wedge \neg\mathsf{sat}_{\psi_2}(s) \to$$
$$\mathsf{value}_{\psi,i}(s) = \frac{1}{2}\Big(\mathsf{value}_{\psi,i-1}(\mathsf{left}(s)) + \mathsf{value}_{\psi,i-1}(\mathsf{right}(s))\Big),$$
$$\forall s \cdot \neg\mathsf{exists}(s) \to \mathsf{value}_{\psi,i}(s) = \frac{1}{2} \cdot \Big(\mathsf{value}_{\psi,i}(\mathsf{left}(s)) + \mathsf{value}_{\psi,i}(\mathsf{right}(s))\Big).$$

The first two constraints deal with the case where $\psi_2$ is true at a real state, and the case where $\psi_1$ and $\psi_2$ are both false at a real state, respectively. The third constraint deals with real states at which $\psi_1$ is true, and $\psi_2$ is false. In this case, the probability that $\psi_1 \, \mathcal{U}^{\leq i} \, \psi_2$ holds at $s$ is the same as the probability that $\psi_1 \, \mathcal{U}^{\leq i-1} \, \psi_2$ holds in the next real state. Hence, the third constraint takes the average of $\mathsf{value}_{\psi,i-1}$ over the successors of $s$. Finally, the fourth constraint deals with the hidden states. Since moving through a hidden state does not count towards the step bound of the until, we take an average of $\mathsf{value}_{\psi,i}$ over the successors for the hidden states.

Having defined the function $\mathsf{value}_{\psi,i}$ for all $i$ in the range $0 \leq i \leq n$, we can now define the function $\mathsf{sat}_\psi$ by comparing the value given by $\mathsf{value}_{\psi,n}$ with the bound $\lambda$.

$$\forall s \cdot \mathsf{exists}(s) \wedge \mathsf{sat}_\psi(s) \leftrightarrow \mathsf{value}_{\psi,n}(s) \bowtie \lambda.$$

**The formula $\phi$**

At this point, we have introduced constraints over $\mathsf{sat}_\psi$ for every sub-formula of the input formula $\phi$. Our final task is to ensure that $\phi$ itself holds at some real state in the model. To do this, we arbitrarily pick State 1, and we require that State 1 is real, and that $\phi$ holds at State 1. Hence, to complete our reduction, we add the constraint:

$\mathsf{exists}(1) \wedge \mathsf{sat}_\phi(1).$

## 5 Implementation and results

### 5.1 Implementation

In this section we describe an implementation of the reduction given in Section 4. In fact, we implement a slightly simpler version of the reduction. In particular, it would obviously be inefficient to produce the function $\mathsf{sat}_\psi$ for the case where $\psi$ is of the form $\neg\psi'$, $\psi_1 \wedge \psi_2$, or $a$ for some atomic proposition $a \in AP$. Instead, we carry out the reduction as normal, and then we iteratively apply the identities given in (1), (2), and (3). For example, we replace all instances of $\mathsf{sat}_{\psi_1 \wedge \psi_2}$ with $\mathsf{sat}_{\psi_1}(s) \wedge \mathsf{sat}_{\psi_2}(s)$. We iterate this procedure until none of the three identities can be applied.

Our implementation consists of a parser that reads PCTL formulas, performs the reduction to SMT, and then outputs the system of SMT constraints. To solve the system of constraints, we experimented with several prominent SMT solvers, and we found that the Yices [17] solver was by far the fastest for our inputs. Hence, all the results described in this section were obtained using Yices-1.0.32 on a machine with a 2.66 GHz Core i7 processor and 4 GB of RAM. The implementation as well as the examples we report on are available for download at `http://www.csc.liv.ac.uk/~john/pctl-smt.tar.bz2`.

Our initial intention was to test our techniques against PCTL formulas from the literature. However, after attempting to find such formulas, we ran into a problem: the PCTL formulas that we found in the literature are not interesting from the perspective of satisfiability. Formulas that are given as examples in model checking papers are often very simple, because the authors are usually interested in the performance when measured in the size of the system. This meant that most of the formulas that we found had extremely simple satisfying models. For example, all of the formulas that appear in [14] have 1 state satisfying models, and as we shall see, these instances are not challenging for our implementation. The same problem occurs for all other examples that we found in the literature. Therefore, in the following subsections, we construct two scalable PCTL formulas that can be used to measure the performance of our implementation.

### 5.2 The lossy channel example

In this section, we test how well our implementation can construct systems. We define a formula $\mathsf{channel}_u$ that represents a lossy channel with $u$ users. For each $i$ in the range $1 \leq i \leq u$, there is an atomic proposition $\mathsf{send}_i$, which indicates that user $i$ wishes to send a message, and an atomic proposition $\mathsf{deliver}_i$, which indicates that the message belonging to

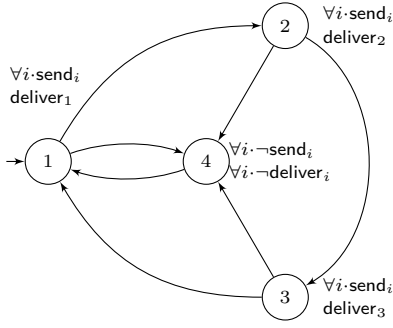

| Users | Time (s) |
|---|---|
| 2 | 0.063 |
| 3 | 0.765 |
| 4 | 12.879 |
| 5 | 18.172 |
| 6 | 8784.490 |

**Figure 4** Experimental results for $\mathsf{channel}_u$

**Figure 3** A model for $\mathsf{channel}_3$

user $i$ has been delivered. The formula $\mathsf{channel}_u$ is defined to be:

$$\mathsf{channel}_{1,u} := \mathbb{P}_{\geq 0.1}(\bigcirc \bigwedge_{1 \leq i \leq u} \neg \mathsf{deliver}_i)$$

$$\mathsf{channel}_{2,u} := \bigwedge_{1 \leq i \leq u} \mathbb{P}_{=0.5}(\bigcirc \mathsf{send}_i)$$

$$\mathsf{channel}_{3,u} := \bigwedge_{1 \leq i \leq u} \Big(\mathsf{send}_i \to \mathbb{P}_{=1}(\top \,\mathcal{U}\, \mathsf{deliver}_i)\Big)$$

$$\mathsf{channel}_{4,u} := \bigwedge_{1 \leq i \leq u} (\mathsf{deliver}_i \to \bigwedge_{1 \leq k \leq u, k \neq i} \neg \mathsf{deliver}_k)$$

$$\mathsf{channel}_u := \mathbb{P}_{=1}\Big(\square \,(\mathsf{channel}_{1,u} \wedge \mathsf{channel}_{2,u} \wedge \mathsf{channel}_{3,u} \wedge \mathsf{channel}_{4,u})\Big).$$

The formula $\mathsf{channel}_{1,u}$ specifies that the channel is lossy: specifically that at least ten percent of the time the channel should not deliver a message at all. The formula $\mathsf{channel}_{2,u}$ specifies that the users should actually use the channel. It states that, in each step, there is a fifty percent chance that each user attempts to send a message. The formula $\mathsf{channel}_{3,u}$ states that, once a message has been sent, it should eventually be delivered. Finally, the formula $\mathsf{channel}_{4,u}$ states that the channel may only deliver one message in each step.

The formula $\mathsf{channel}_u$ has a model with $u+1$ states. For example, in Figure 3, we show the structure of a model for $\mathsf{channel}_3$. The atomic propositions $\mathsf{send}_1$, $\mathsf{send}_2$, and $\mathsf{send}_3$, are true in all states but state 4. We also have that $\mathsf{deliver}_1$ is true in state 1, that $\mathsf{deliver}_2$ is true in state 2, and that $\mathsf{deliver}_3$ is true in state 3. Other than these exceptions, the $\mathsf{deliver}_i$ atomic propositions are false at all other states.

It is also not difficult to see that the formula $\mathsf{channel}_u$ cannot have a model with fewer than $u+1$ states. This is because, with probability 1, each of the $\mathsf{send}_i$ atomic propositions must become true. Once this has occurred, we require at least $u+1$ distinct states: since two messages cannot be delivered at the same time, we require at least one state for each $\mathsf{deliver}_i$, and since the channel is lossy, we require at least one state in which no $\mathsf{deliver}_i$ is true.

In the table in Figure 4, we show experimental results for the formula $\mathsf{channel}_u$. In each case, we asked our implementation to find a model with $u+1$ states for the formula $\mathsf{channel}_u$. These results show that our implementation can construct systems with a small number of states very quickly. However, the exponential nature of the problem catches up to us quite quickly, and checking for the existence of a model with 7 states for the formula channel$_6$ already takes more than two hours. The SMT solver was not able to solve the constraint system for channel$_7$ within a reasonable amount of time.

## 5.3 A lossy channel with bugs

In this section we study the ability of our implementation to find bugs in specifications. We define an extension to our lossy channel that allows the system to go down, and we will require that, if the system does go down, then a number of recovery steps will be taken in order to restore service. Unfortunately, our formula will have a bug, and our goal is to find out how well our implementation can find this bug.

Our formula will be called broken$_{u,r}$, which represents a lossy channel with $u$ users, and which can recover from an error in $r$ steps. In addition to the atomic propositions used by channel$_u$, we add several new atomic propositions. The atomic proposition up indicates whether the network is up or down. For each $j$ in the range $1 \leq j < r$, there is an atomic proposition recover$_j$, that indicates that the network is in the $j$th step of the recovery procedure. We will require that each of the recovery propositions must be true before the system can come back up.

Our formula will reuse the formulas channel$_{1,u}$ and channel$_{4,u}$ from the previous section. However, we replace the other two formulas with the following:

$$\mathsf{broken}_{2,u,r} := \bigwedge_{1 \leq i \leq u} \Big(\mathsf{up} \land \bigwedge_{1 \leq j \leq r} \neg\mathsf{recover}_j\Big) \to \mathbb{P}_{=0.5}(\bigcirc \mathsf{send}_i)$$

$$\mathsf{broken}_{3,u,r} := \bigwedge_{1 \leq i \leq u} \Big(\mathsf{up} \land \mathsf{send}_i \to \mathbb{P}_{=1}(\top \,\mathcal{U}\, \mathsf{deliver}_i)\Big)$$

$$\mathsf{broken}_{5,u,r} := \bigwedge_{1 \leq i \leq u} \Big(\neg\mathsf{up} \to \bigwedge_{1 \leq k \leq u, k \neq i} \neg\mathsf{deliver}_k\Big)$$

The formula broken$_{2,u,r}$ specifies that, if the channel is up, and not in a recovery state, then users should be able to send messages. The formula broken$_{3,u,r}$ specifies that, if a user sends a message while the network is up, then that message should be delivered. Finally, the formula broken$_{5,u,r}$ specifies that, if the network is not up, then no messages can be delivered.

In addition to these formulas, we also specify the recovery procedure. We define:

$$\mathsf{broken}^0_r := \begin{cases} \neg\mathsf{up} \to \mathbb{P}_{\geq 0.99}(\bigcirc \mathsf{up}) & \text{if } r = 1, \\ \neg\mathsf{up} \to \mathbb{P}_{\geq 0.99}(\bigcirc \mathsf{recover}_1) & \text{otherwise,} \end{cases}$$

and, for all $j$ in the range $1 \leq j \leq r$, we define:

$$\mathsf{broken}^0_j := \begin{cases} \mathsf{recover}_j \to \mathbb{P}_{=1}(\bigcirc \mathsf{recover}_{j+1} \land \bigwedge_{1 \leq k \leq j} \neg\mathsf{recover}_k) & \text{if } j < r, \\ \mathsf{recover}_j \to \mathbb{P}_{=1}(\bigcirc \mathsf{up} \land \bigwedge_{1 \leq k \leq r} \neg\mathsf{recover}_k) & \text{if } j = r. \end{cases}$$

If $r = 1$, then these formulas specify that the system should recover in one step after it has gone down. For other values of $r$, these formulas specify that the system should pass through each of the recovery states before the channel comes back up. We can now specify the full formula:

$$\mathsf{broken}_{u,r} := \mathbb{P}_{=1}\Big(\Box\,(\mathsf{channel}_{1,u} \land \mathsf{broken}_{2,u,r}$$
$$\land\, \mathsf{broken}_{3,u,r} \land \mathsf{channel}_{4,u} \land \mathsf{broken}_{5,u,r} \land \bigwedge_{1 \leq j \leq r} \mathsf{broken}^j_r\Big)\Big).$$

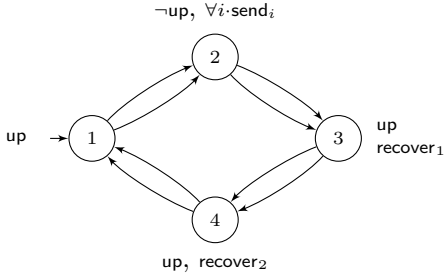

**Figure 5** A model for broken$_{u,3}$

| Users | Time (s) | | | |
|---|---|---|---|---|
| | $r=1$ | $r=2$ | $r=3$ | $r=4$ |
| 10 | 0.192 | 0.376 | 13.852 | 89.908 |
| 20 | 0.609 | 1.643 | 2.666 | 72.341 |
| 30 | 0.869 | 2.723 | 58.825 | 278.150 |
| 40 | 1.139 | 2.153 | 23.023 | 167.234 |
| 50 | 1.886 | 10.695 | 55.375 | 336.153 |
| 60 | 4.510 | 9.624 | 64.603 | 502.790 |
| 70 | 6.912 | 11.622 | 54.162 | 216.548 |
| 80 | 7.672 | 40.855 | 55.568 | 370.268 |
| 90 | 4.457 | 46.538 | 151.892 | 565.718 |
| 100 | 12.111 | 47.440 | 231.619 | 1927.258 |

**Figure 6** Experimental results for broken$_{u,r}$

Unfortunately, this formula contains a bug: if the system immediately goes down after being up, then no messages are ever delivered. Figure 5 shows the structure of a model that satisfies broken$_{u,3}$ for all $u$. We have that up is satisfied in all states except 2, that recover$_1$ is satisfied in state 3, and that recover$_2$ is satisfied in state 4. The propositions deliver$_i$ are not satisfied in any state, and the propositions send$_i$ are only satisfied in state 2.

We asked our implementation to solve broken$_{u,r}$ for varying values of $u$ and $r$. The results are displayed in Figure 6. These results show a similar scaling with respect to $r$ as was found for the previous example: as the size of the smallest model increases, our performance gets progressively worse, and we were unable to obtain results for $r = 5$ within a reasonable amount of time. However, these results show that our implementation scales well with respect to the number of users. This indicates that, while the running time of the procedure depends strongly on the size of the minimal model, there is a much smaller dependence on the size of the PCTL formula. Indeed, the formula broken$_{100,r}$ contains over 300 temporal operators. It is for this reason that we claim that our techniques are particularly suited for sanity checking, because this application requires us to construct a small model for a complex formula. These results show that our procedure can handle such situations.

## 6   Conclusion

In this paper, we have introduced the *bounded satisfiability* problem, which is a simplification of the satisfiability problem for PCTL, that restricts consideration to models that can be implemented. As was expected, bounded satisfiability is NP-complete in the minimal output. To offset this negative result, we provided a reduction from bounded satisfiability to an SMT problem, because in practice SMT solvers can often answer large queries.

Our experimental results allow for two interpretations. Our first set of experimental results shows that the difficulty of finding a solution to the system of SMT constraints depends strongly on the size of the minimal model. Hence, we consider it unlikely that these techniques will be able to construct the large systems that would be useful in practice. On the other hand, our second set of benchmarks showed that the running time of the SMT solver does not depend strongly on the size of the PCTL formula. Indeed, we were able to construct systems that satisfy formulas with hundreds of temporal operators. It would seem that, while our techniques are not able to construct models that are large enough be useful in practice, they are able to handle the large specifications that may appear in this setting. This motivates the idea of sanity checking, where a system designer wishes to ensure that there are no errors in a specification that could lead to a small satisfying model. Our results indicate that our techniques are capable of providing a sanity checking procedure.

## A  Proof of Proposition 8

**Proof.** Inclusion in NP has been shown through the reduction to SMT (with linear arithmetic and uninterpreted functions. To establish that the problem is NP-hard, we will reduce from 3SAT. Let $\psi$ be an instance of 3SAT over the set of propositions $P$. We define a PCTL formula $\phi$ over the set of atomic propositions $AP = P \cup \{p_1, p_2, \ldots, p_{\lceil \log_2(b+1) \rceil}\}$. The extra atomic propositions $p_i$ will be used as bit strings to represent numbers. For each $i$ in the range $1 \le i \le b+1$, we define $\mathsf{truth}_i$ to give the expansion of $i$ in binary using these propositions. For example, if $b + 1 = 3$, then we will have the atomic propositions $p_1$ and $p_2$. Therefore, we set

$$\mathsf{truth}_0 := \neg p_1 \wedge \neg p_2 \qquad\qquad \mathsf{truth}_1 := p_1 \wedge \neg p_2$$
$$\mathsf{truth}_2 := \neg p_1 \wedge p_2 \qquad\qquad \mathsf{truth}_3 := p_1 \wedge p_2$$

The definition of $\phi$ is as follows. For each $i$ in the range $1 \le i \le b-1$ we define:

$$\phi_i := \mathbb{P}_{=1}\Big(\Box \left(\mathsf{truth}_i \to \mathbb{P}_{=1}\big(\bigcirc \mathsf{truth}_{i+1}\big)\right)\Big),$$

which says that, if we are in a state that satisfies $\mathsf{truth}_i$, then we must move to a state that satisfies $\mathsf{truth}_{i+1}$. We also define:

$$\phi_b := \mathbb{P}_{=1}\Big(\Box \left(\mathsf{truth}_b \to \psi \vee \mathbb{P}_{=1}(\bigcirc \mathsf{truth}_{b+1})\right)\Big),$$

which says that, if we are in a state that satisfies $\mathsf{truth}_b$, then either the state must satisfy the boolean formula $\psi$, or the next state must satisfy $\mathsf{truth}_{b+1}$. Finally, our formula is defined to be the conjunction of these formulas:

$$\phi := \mathsf{truth}_1 \wedge \bigwedge_{1 \le i \le b} \phi_i.$$

We argue that the PCTL formula $\phi$ has a model with $b$ states if and only if the boolean formula $\psi$ has a satisfying assignment. It is clear that $\phi$ does not have a model with $b-1$ states, because our formula forces the model to have at least one state for each $\mathsf{truth}_i$ in the range $1 \le i \le b$, and if $\mathsf{truth}_i$ holds at a state $s$, then $\mathsf{truth}_j$, with $j \ne i$, cannot hold at $s$. If the boolean formula $\psi$ has a satisfying assignment, then the formula does have a model with $b$ states, since we can satisfy the PCTL sub-formula $\phi_b$ with a single state that satisfies $\mathsf{truth}_b$ and $\psi$. On the other hand, if the boolean formula does not have a satisfying assignment, then the PCTL sub-formula $\phi_b$ forces the model to contain a state satisfying $\mathsf{truth}_{b+1}$. Therefore, the PCTL formula $\phi$ has a model with $b$ states if and only if the boolean formula $\psi$ has a satisfying assignment. ◀

## B  Proof of Proposition 9

**Proof.** We will again provide a reduction from 3SAT. Let $\psi$ be an instance of 3SAT over the set of propositions $P$. We define a PCTL formula $\phi$ over the set of atomic propositions $AP = P \cup \{p_1, p_2, \ldots, p_{\lceil \log_2(|\psi|) \rceil}\} \cup \{q_1, q_2, \ldots, q_{|\psi|-1}\}$. The propositions $p_i$ serve the same purpose as they did in the proof of Proposition 8, and we will use the same definition of $\mathsf{truth}_i$ as we did in that proposition.

We begin by defining a formula that forces the construction of a full binary tree of depth $|\psi|$. We begin with by defining the formula:

$$\phi_i^{\mathrm{split}} := \mathbb{P}_{=0.5}\Big(\bigcirc \mathbb{P}_{=1}(\Box\, q_i)\Big) \wedge \mathbb{P}_{=0.5}\Big(\bigcirc \mathbb{P}_{=1}(\Box\, \neg q_i)\Big),$$

which states that there should be a probability of 0.5 that $q_i$ should hold forever in the next state, and there should be a probability of 0.5 that $\neg q_i$ should hold forever in the next state. This formula will be used to force our model to branch. For each $i$ in the range $1 \leq i \leq |\psi| - 1$, we define the PCTL formula:

$$\phi_i := \mathbb{P}_{=1}\Big(\Box \left(\mathsf{truth}_i \to \mathbb{P}_{=1}(\bigcirc \mathsf{truth}_{i+1}) \wedge \phi_i^{\mathrm{split}}\right)\Big),$$

which states that, if $\mathsf{truth}_i$ holds at a state, then $\mathsf{truth}_{i+1}$ must hold in the next state, and requires that $\phi_i^{\mathrm{split}}$ holds. Finally we define our formula:

$$\phi := \psi \vee (\mathsf{truth}_1 \wedge \bigwedge_{1 \leq i \leq |\psi|-1} \phi_i).$$

It is obvious that if the boolean formula $\psi$ has a satisfying assignment, then the PCTL formula $\phi$ has a model with a single state. We will prove that, if the boolean formula $\psi$ does not have a satisfying assignment, then the smallest model for the PCTL formula $\phi$ has $2^{|\psi|} - 1$ states.

We will prove this claim by induction. For each $i$ in the range $1 \leq i \leq |\psi| - 1$, we define:

$$\phi'_i := \mathsf{truth}_i \wedge \bigwedge_{i \leq j \leq |\psi|-1} \phi_i.$$

Our inductive hypothesis will be that the minimal model for $\phi'_i$ has $2^{|\psi|-i+1} - 1$ states.

For the base case, we consider the formula $\phi'_{|\psi|-1}$. Clearly this formula can be satisfied with a $2^{|\psi|-(|\psi|-1)+1} - 1 = 3$ state model, which consists of a starting state with two successors. The starting state satisfies $\mathsf{truth}_{|\psi|-1}$, and both successors satisfy $\mathsf{truth}_{|\psi|}$. Moreover, one successor satisfies $q_{|\psi|-1}$ and the other satisfies $\neg q_{|\psi|-1}$. It is also fairly clear that $\phi'_{|\psi|-1}$ does not have a two state model, since there must be a state that satisfies $\mathsf{truth}_{|\psi|-1}$, and the formula $\phi_{|\psi|-1}^{\mathrm{split}}$ forces this state to have two distinct successors.

We now prove the inductive step. Suppose that the inductive hypothesis has been proved for $\phi'_{i+1}$ for some $i$. We will prove the inductive hypothesis for the formula $\phi'_i$. It is clear that we can construct a model of $\phi'_i$ of size $2^{|\psi|-i+1} - 1$ by creating a starting state $s$ that satisfies $\mathsf{truth}_i$, and adding two minimal models for $\phi'_{i+1}$ as successors to $s$: one model in which $q_i$ holds in every state, and one model in which $\neg q_i$ holds in every state. The total number of states used by this model is:

$$2 \cdot (2^{|\psi|-i} - 1) + 1 = 2^{|\psi|-i+1} - 1.$$

We must now prove that this is a minimal model. We will do so by contradiction. Suppose that $\phi'_i$ has a model $\mathcal{M}$ with fewer than $2^{|\psi|-i+1} - 1$ states. Let $s$ be the initial state of $\mathcal{M}$. The formula $\phi_i^{\mathrm{split}}$ ensures that the state $s$ has at least two successors: it must have a successor $s_1$ that satisfies $q_i$, and it must have a successor $s_2$ that satisfies $\neg q_i$. Moreover, the set of states reachable from $s_1$ must be disjoint from the set of states reachable from $s_2$. Finally, all successors of $s$ must satisfy $\phi'_i$. Hence, we have two disjoint models for the formula $\phi'_i$. Let us denote these models as $\mathcal{M}_1$ and $\mathcal{M}_2$. From our assumption on the size of $\mathcal{M}$, we can derive the following bound for the size of these models:

$$|\mathcal{M}_1| + |\mathcal{M}_2| = |\mathcal{M}| - 1 < 2^{|\psi|-i+1} - 2 = 2 \cdot (2^{|\psi|-i} - 1).$$

Hence, we either have $|\mathcal{M}_1| < 2^{|\psi|-i} - 1$, or we have $|\mathcal{M}_2| < 2^{|\psi|-i} - 1$, and either case contradicts the inductive hypothesis.

To complete the proof, it suffices to note that every model that satisfies $\phi$, either satisfies $\psi$, or satisfies $\phi'_1$. Therefore, if $\psi$ does have a satisfying assignment, then the size of the minimal model is 1. If $\psi$ does not have a satisfying assignment, then the size of the minimal model is $2^{|\psi|} - 1$ states. Since the size of $\phi$ is polynomial in $|\psi|$, we can conclude that it is NP-hard to approximate the size of the minimal model for a PCTL formula to within a factor polynomial in $|\phi|$. ◂